\documentclass[prl,preprintnumbers,tightenlines,twocolumn,superscriptaddress]{revtex4-1}

\usepackage{amsmath}
\usepackage{amssymb}
\usepackage{graphicx}
\usepackage{color}
\usepackage[utf8]{inputenc}

\newcommand{\bra}[1]{\langle #1 |}

\newcommand{\ket}[1]{| #1 \rangle}

\newcommand{\E}[1][\empty]{
  \ifthenelse{\equal{#1}{\empty}}
    {\mathbb{E}}
    {\mathbb{E}\left( #1 \right)}
}
\renewcommand{\exp}[1][\empty]{
  \ifthenelse{\equal{#1}{\empty}}
    {\mathrm{exp}}
    {\mathrm{e}^{#1}}
}

\newcommand{\psit}[1][\empty]{%
  \ifthenelse{\equal{#1}{\empty}}
    {\psi_t}
    {\psi_t^{(#1)}}
}

\newcommand{\npsit}[1][\empty]{%
  \ifthenelse{\equal{#1}{\empty}}
    {\tilde\psi_t}
    {\tilde\psi_t^{(#1)}}
}

\newcommand{\SI}{Supplemental Material}

\newcommand{\up}{\uparrow}
\newcommand{\down}{\downarrow}

\usepackage{ulem}  %unter- (\uline{important}) und durchstreichen (\sout{wrong})
\newcommand{\oalex}[1]{{\color{blue}{}}}% \sout{#1}}}

\definecolor{olive}{RGB}{107,142,35}

\definecolor{orange}{RGB}{255,139,61}

\graphicspath{{{./}}}
\begin{document}
\title{Extended coherently delocalized states in a frozen Rydberg gas}

\author{G. Abumwis}
\affiliation{Max-Planck-Institut f\"ur Physik komplexer Systeme, N\"othnitzer Str.\ 38, 
D-01187 Dresden, Germany }
	
\author{Matthew T. Eiles}
\affiliation{Max-Planck-Institut f\"ur Physik komplexer Systeme, N\"othnitzer Str.\ 38, 
D-01187 Dresden, Germany }

\author{Alexander Eisfeld}
\email{eisfeld@pks.mpg.de}
\affiliation{Max-Planck-Institut f\"ur Physik komplexer Systeme, N\"othnitzer Str.\ 38,
D-01187 Dresden, Germany }

\begin{abstract}
The long-range dipole-dipole interaction can create delocalized states due to the exchange of excitation between Rydberg atoms. We show that even in a random gas many of the single-exciton eigenstates are surprisingly delocalized, composed of roughly one quarter of the participating atoms. We identify two different types of eigenstates: one which stems from strongly-interacting clusters, resulting in localized states, and one which extends over large delocalized networks of atoms. These two types of states can be excited and distinguished by appropriately tuned microwave pulses, and their relative contributions can be modified by the Rydberg blockade and the choice of microwave parameters.

\end{abstract}
\maketitle

Assemblies of cold Rydberg atoms are {ideally suited} to investigate interactions in many-particle systems.They possess many readily tunable properties and can, in many circumstances, be treated with essential state Hamiltonians, easing theoretical interpretation \cite{ Browaeys2016a,GallagherBook,Lukin2001,Browaeys2016,Saffman,low2012experimental}. Although in recent years several groups have successfully created well-defined and reproducible structures of Rydberg atoms \cite{Birkl2019,Lukin2017,Gross2019,Browaeys2018}, the most common experimental scenario is a frozen Rydberg gas \cite{Pillet1998,GallagherFroze1998}. In such an environment, the Rydberg atoms are distributed randomly and are immobile over typical experimental timescales due to the ultracold temperature \cite{Robicheaux2004}.

A particularly clear example of collective states of a randomgas with $N$ {Rydberg} atoms is given by the following scenario: {we consider} two states per {Rydberg} atom, denoted {$\up\equiv\nu \mathit{s}$ and $\down\equiv\nu \mathit{p}$, with energies $\epsilon_\up$ and $\epsilon_\down$.
Here, $\nu$ is the principal quantum number while {$s$ and $p$ indicate the orbital angular momentum.}
  We focus on the single-exciton sector of the full Hamiltonian, which is spanned by the degenerate states $\ket{n} = \ket{\down\down\dots\up\dots \down}$ which have energy $\epsilon_n = \epsilon_\up^{(n)}+\sum_{j\ne n}\epsilon_\down^{(j)}$. This notation implies a labeling scheme for the atoms where the sole $\up$ excitation lies at atom $n$. Because of resonant dipole-dipole interactions \cite{Fr30_198_a} the states $\ket{n}$ are not energy eigenstates, but the eigenstates {instead have} the form 
\begin{equation}
\ket{\psi_\ell}=\sum_n c_n^{(\ell)} \ket{n}.
\end{equation}
The coefficients $c_n^{(\ell)}$ determine the extent to which $\ket{\psi_\ell}$ is coherently delocalized. 
Delocalization can be a challenging concept to quantify since it is a property of the wave function itself, and not a simple observable. Several complementary measures can be used to extract the most relevant information \cite{KrMa93_1469_,GoLiZh16_59_}; two standard ones are the ``inverse paticipation ratio''  \cite{KrMa93_1469_}  and the coherence \cite{BaCrPl14_140401_},
\begin{equation}
\label{eq:coherence}
%\mathcal{C_\ell}=\sum_{n}\sum_{m\ne m} |\rho^{(\ell)}_{nm}| ,
\mathcal{C_\ell}=\sum_{n}\sum_{m\ne m} \left|(c^{(\ell)}_n)^* c^{(\ell)}_m\right|.
\end{equation}
   We focus on coherence since it has a more intuitive intepretation. As a rule of thumb, its value roughly corresponds to the number of atoms coherently sharing the $\up$ excitation. $\mathcal{C}=1$ corresponds exactly to a dimer state {and for a equally distributed state ($c_n=1/\sqrt{N}$) its value is $N-1$. We provide further examples in the \SI.}

\begin{figure}[bp]
\includegraphics[width=1.\columnwidth]{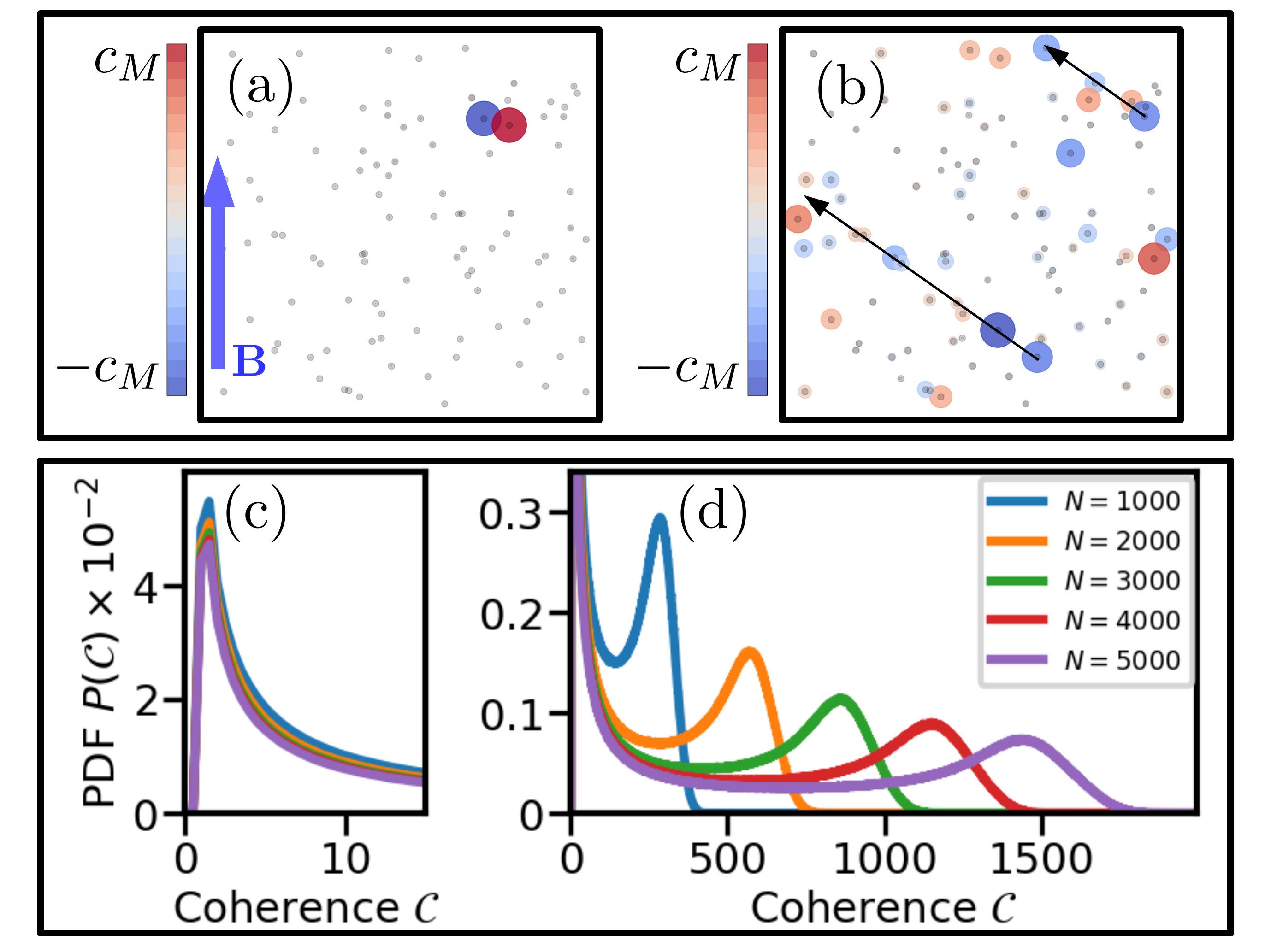}
\caption{\label{fig:intro} (a) and (b): Two different eigenstates of the same realization (2D case).  The circle size and color represents the $\up$ amplitude at each site. The magnetic field axis (blue arrow) and magic angle of the dipole-dipole interaction (black arrows) are discussed in the text. For each state $c_M = \max_n c_n^{(\ell)}$.  (c) and (d): The probability to find states with coherence $\mathcal{C}$ of a 3D random Rydberg gas with various number of Rydberg atoms $N$. Panel (c) highlights the low coherence and (d) the high coherence regions. We averaged over $10^4$ random realizations. Note the different scales of the $y$ axes. }

\end{figure} 
It is well-known that dimer states having $c_n^{(\ell)}\ne 0$ at just two atoms form {because in }a random Rydberg gas there exist pairs of atoms with interparticle separations far {smaller} than the mean nearest-neighbor distance. Fig.~\ref{fig:intro}a shows one of these dimers in, for pictorial clarity, a two-dimensional gas.  {The two atoms in this dimer }interact strongly and decouple energetically from the rest of the gas, and exhibit a range of fascinating behavior and dynamics \cite{Plasma1,Plasma2,Deiglmayr2016,Cote2002,Gross2019}. There are numerous studies which have investigated the rich physics of the full eigenenergy spectrum  \cite{RoHeTo04_042703_,ScWeBu14_63415_,Goetschy,Goetschy2}.
What is far less understood is the extent to which coherently delocalized eigenstates can develop given the random nature of the gas.  Because of the clustering properties of a random gas and the energetic decoupling of strongly interacting clusters of atoms (dimers, trimers, etc.), one could surmise that the gas fragments into a hierarchy of clusters with corresponding eigenstates that remain small relative to the total gas size.  In contrast to this hypothesis, the state in Fig.~\ref{fig:intro}(b) exhibits remarkably large delocalization over many atoms. Panel (d) shows that this delocalization is not unique to this state. Delocalization is quantified by the coherence $\mathcal{C}$ (defined in Eq.~\eqref{eq:coherence}). 
Clearly, states involving around one third of the atoms are very common.

Before we investigate the properties of these states further, we first provide more details about the physical system and our theoretical modeling. 
A possible way to investigate the single-exciton eigenstates $\ket{\psi_\ell}$ is via microwave transitions from the state $G = \ket{\down \down...\down}$ with all atoms initially in the $\down$ state. 
The dimension of $G$  determines the number of Rydberg atoms involved ($N$). 
In a typical scenario, roughly 1\% of the ground state atoms in a gas can be promoted to the $\down$ Rydberg state, and so the Rydberg density $n$ can easily range from $10^{7}-10^{12}$cm$^{-3}$  \cite{low2012experimental}.  For  ultracold gas dimensions of $V\sim(200\mu\text{m})^3$, this process results in $N\approx1000$ Rydberg atoms. In our simulations we place $N$ Rydberg atoms within a cube following a uniform distribution of positions $\vec R_n$. Although realistic atomic clouds do not have truly uniformly distributed particles, we {show in the \SI{} that particles distributed according to a Gaussian distribution have qualitatively  similar coherence properties.} 

In general, each state $\ket{n}$ possesses degenerate magnetic quantum number sublevels and the interaction has a tensorial form  \cite{PaTaCl11_22704_,MoeWueAt11_184011_,RoHeTo04_042703_}. We avoid this complication by applying a $\sim10$G magnetic field to isolate the $m_l=0$ subspace via the Zeeman shift of $1.4\,$MHz/G {as often done in experiment Ref.~\cite{BaLaRa15_113002_}}. Then, the relevant Hamiltonian is
\begin{align}
\label{eq:ham}
H&= \sum_{n=1}^N \epsilon_n \ket{n}\bra{n}+\sum_n \sum_{m\ne n} V_{nm}(\vec{R}_n,\vec{R}_m)\ket{n}\bra{m},
\end{align}
with
\begin{equation}
\label{eq:dipdip}
V_{nm}(\vec{R}_n,\vec{R}_m)= \frac{1}{3}\frac {\mu^2}{|\vec R_n - \vec R_m|^3}(1-3\cos^2\theta),
\end{equation}
where $\theta$ is the relative angle between $\vec R_{n}-\vec R_m$ and $\vec B$.
The transition dipole between $\up$ and $\down$ states is denoted by $\mu$. 
Eq. \ref{eq:dipdip} is modified when retardation effects are relevant, but these can be neglected for our system size of a few millimeters and transition frequencies $\omega_{ps}=\epsilon_p- \epsilon_s$ of several GHz.

  We now examine the relevant properties of the eigenstates of the Hamiltonian (Eq. \ref{eq:ham}).
{  These eigenstates are obtained by numerically diagonalizing the Hamiltonian for a large number of atomic arrangements \footnote{Details about the numerical procedure are provided in the \SI.}.}
 As seen in {Fig.~\ref{fig:intro}c}, there are no states with coherence $\mathcal{C}$ smaller than one, implying that it is impossible to excite individual atoms in the gas. 
 Following the sudden onset at $\mathcal{C} = 1$, i.e. the appearance of dimers, the coherence probability rapidly decreases at a rate nearly independent of $N$, before leveling off and continuing at a finite value into a very long tail ({Fig.~\ref{fig:intro}d}). 
 The tail extends to coherence values around one-third of $N$, and even increases to form a broad peak at large $\mathcal{C}$. 

\begin{figure}[t]
\includegraphics[width=\columnwidth]{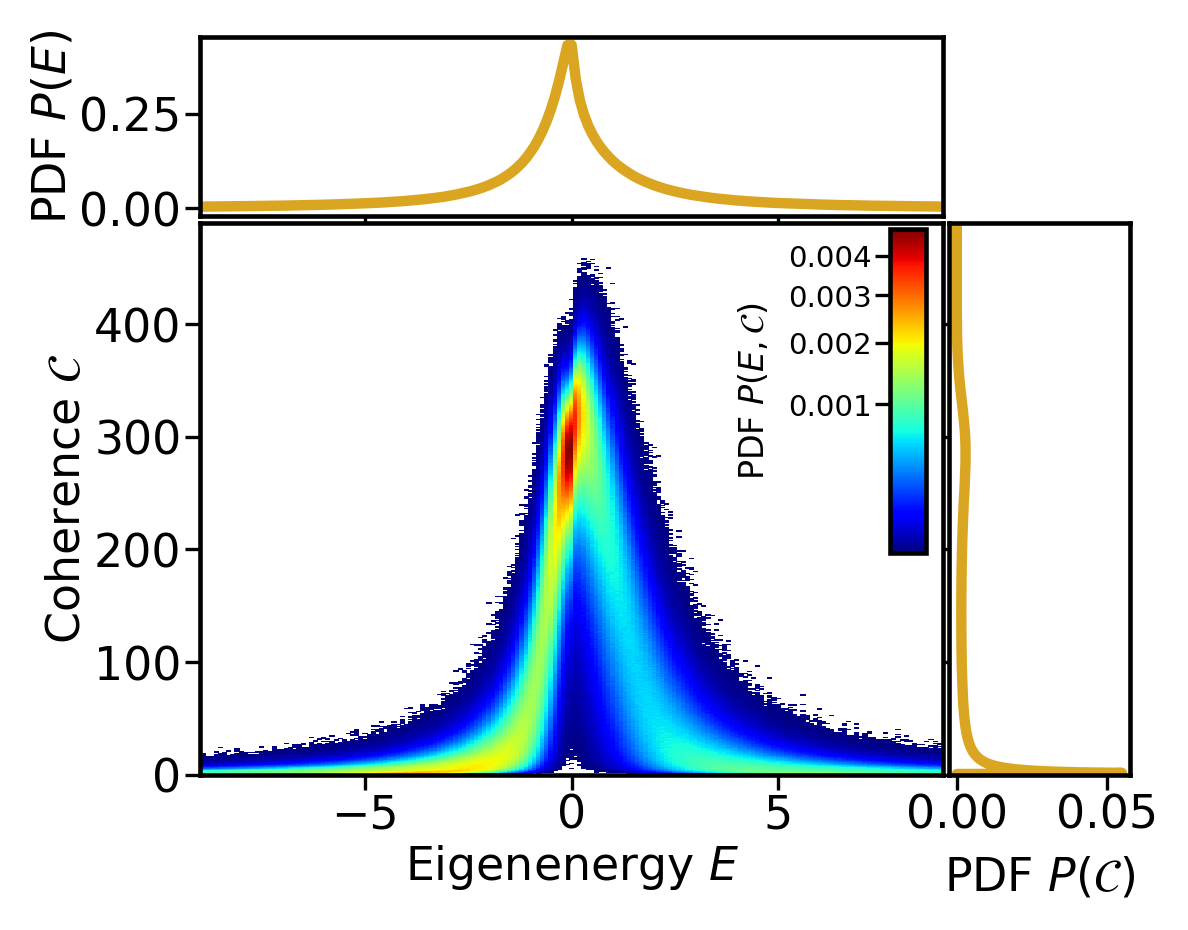}
\caption{The distribution (probability density function) of states having coherence $\mathcal{C}$ and eigenenergy $E$ for $N=1000$. 
The zero of energy is the energy $\epsilon_n$ of non-interacting atoms. As the unit of energy we use $V_0\equiv\frac{4 \pi}{9}\mu^2 n$, which comes from evaluating Eq.~(\ref{eq:dipdip}) for $\theta=0$ at the Wigner-Seitz radius $a=(\frac{3}{4\pi n})^{1/3}$. It corresponds to the typical energy scale of a Rydberg gas with density $n$.
The marginal distributions are plotted on the top (corresponds to the density of states) and side (cf.~Fig.~\ref{fig:intro}c,d).  \label{fig:coherence_eigen_3D}}
\end{figure}

 To gain more insight into this coherence distribution we investigate the correlation between eigenenergy and delocalization. Fig.~\ref{fig:coherence_eigen_3D}  displays the probability to find a state with a given eigenenergy and coherence. This distribution clearly reveals that the low coherence peak in Fig.~\ref{fig:intro} is associated with large energy shifts; the energy tails (not shown at this scale) are almost exclusively dimer states with $\mathcal{C} = 1$.  Since the probability to find small clusters of atoms is independent of $N$, so is the coherence probability over this range, as confirmed by Fig.~\ref{fig:intro}c. In contrast, states with high coherence are strongly associated with states having approximately the mean interaction energy. This suggests that these large decoherences are provided by networks of mutually interacting atoms. 
 
 To better understand what aspects of the interaction $V_{nm}$ are responsible for the appearance of the delocalized states and their distribution, we have varied the 'long-range character' and the angular form of the interaction by choosing different power law exponents $R^{-\alpha}$ and neglecting the angular dependence \cite{ourJPB}. We always find many delocalized states. The overall coherence decreases as the interaction becomes more short-ranged.  We observe also that increased anisotropy in the interaction increases the amount of delocalization. The presence of the anistropy complicates a simple one-to-one correspondence between small interparticle distances and large interactions, and could be a reason why extended networks featuring large coherence are more probable for anisotropic interactions. This is because close pairs at the magic angle where the interaction vanishes, $\theta\approx 54^\circ$, do not interact{. They therefore} become part of extended states rather than dimers;  this is manifested in the angular correlations along the rays visible in Fig.~\ref{fig:intro}b.

To study the delocalized states further, we take advantage of an inherent mechanism to suppress the population of dimers in a random Rydberg gas: the Rydberg blockade \cite{JaCiZo00_2208_,tong2004local,gaetan2009observation,urban2009observation}.
 This has a profound impact on the distribution of atomic positions making up the initial state $\ket G$ because two atoms closer than the blockade radius, $R_B\propto (\nu^{11}/\Omega)^{1/6}$, cannot be simultaneously excited (we ignore the anisotropy that can, depending on the atomic states being considered, be present in the induced van der Waals interaction.)  In the laboratory, varying $\nu$ or the laser bandwidth $\Omega$ can tune the blockade radius over a wide range of values. To crudely incorporate the Rydberg blockade we eliminate, from the initial distribution of Rydberg atom positions, one atom from each pair having a mutual separation less than one $R_B$. 
The Rydberg blockade allows us to relate localization and coherence to the interparticle separations in the gas  \cite{ScWeBu14_63415_}, since it prevents the formation of small clusters of atoms, eliminating eigenstates with small coherence like the one shown in Fig.~\ref{fig:intro}a. 
Indeed, Fig.~\ref{fig:blockade} reveals a sharp loss in the peak at low coherence.   
As the blockade radius increases to the Wigner-Seitz radius the low coherence peak is totally erased, compensated by an increase in the number of highly delocalized states.

After compiling these results together, an explanation of the formation of  delocalized states emerges. 
It is clear that strongly localized states are associated with very strong interactions, and hence with small clusters at favorable orientations for the dipole-dipole anisotropy. 
These clusters decouple from and cease to interact with the rest of the system, leaving behind a residual distribution of atoms which is no longer truly uniformly distributed since it has very few remaining small clusters (the Rydberg blockade exaggerates this  by even more strongly suppressing cluster formation in the initial distribution). 
The remaining atoms left to participate are still randomly arranged, but their spacing is more regular than in a uniform distribution. 
The excitation therefore extends over very many atoms. 
We note that, as most previous effort has been devoted to the eigenvalue statistics of such random systems, rather than their eigenstate properties, this property has to the best of our knowledge only scarcely been noticed \cite{scholak2014spectral,Skipetrov2011,Akulin2014}.

Of course, these coherent delocalized states are only physically relevant if they are robust to noise or disorder. If perturbations on the order of the smallest interactions in the gas could destroy these states, then the delocalization is in some sense trivial and, more crucially, could never be realized experimentally. A sophisticated study of the effects of disorder and decoherence requires a full inclusion of these effects into the evolution of the density matrix, which is beyond the scope of this Letter. Instead, as a crude check of the effects of some of these perturbations, we include diagonal disorder by randomly varying $\epsilon_n$ according to a  uniform distribution, or remove small off-diagonal matrix elements under {a} cutoff threshold (see Ref.~\cite{ourJPB} for more details). 
We express the  width of the distribution (i.e.~the strength of the disorder) and this cutoff threshold in units of $V_0=\frac{4\pi}{9}\mu^2 n$, the interaction strength at the Wigner-Seitz radius.
 Both effects tend to suppress the long-range coherence, but we find that this suppression is not strong in this system: 
 the localization length is only reduced by a factor of around two-thirds even when the disorder strength is on the order of $V_0$ or when interactions up to a tenth of $V_0$ are removed.
  This shows that these states are robust, and furthermore indicates that the interactions between various atoms contained in the delocalized states are still fairly large, which helps to preserve the delocalization under perturbation.

\begin{figure}
\includegraphics[width=1.\columnwidth]{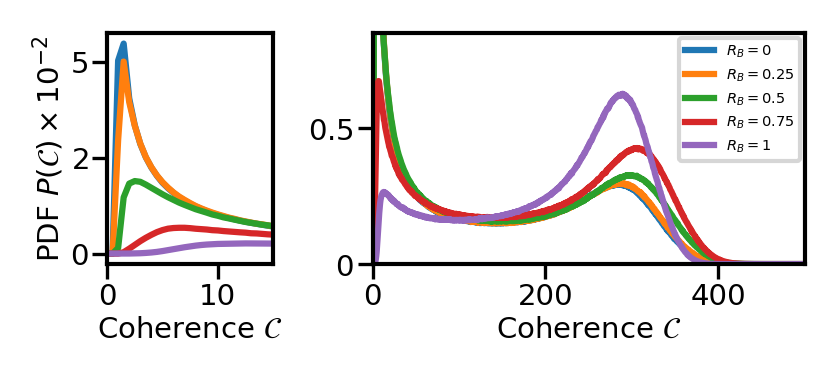}
\caption{\label{fig:blockade}
Coherence probability for several Rydberg blockade radii $R_B$ given in units of the Wigner-Seitz radius $a$.
 The two panels highlight different regions and use different $y$-scales.
For $R_B=0$ the number of Rydberg atoms is $N = 1000$; as a result of the blockade this reduces to $992$ for $R_B = 0.25$, $943$ for $R_B = 0.5$, $834$ for $R_B = 0.75$, and $686$ for $R_B = 1$. $5\times 10^5$ realizations were used.  {The resonant and non-resonant interactions scale rather differently with $\nu$ and density $n$, $V_R\sim \nu^4n$ and $V_{NR}\sim\nu^4n^2$. To keep the resonant interactions to a reasonable level, around $10$MHz, while realizing the Rydberg blockade with blockade radius $a$ for a typical laser bandwidth of 1MHz, would require relatively high Rydberg states $(\sim 100)$ and low densities $(\sim10^7)$cm$^{-3}$.}
}
\end{figure}

\begin{figure}
\includegraphics[width=\columnwidth]{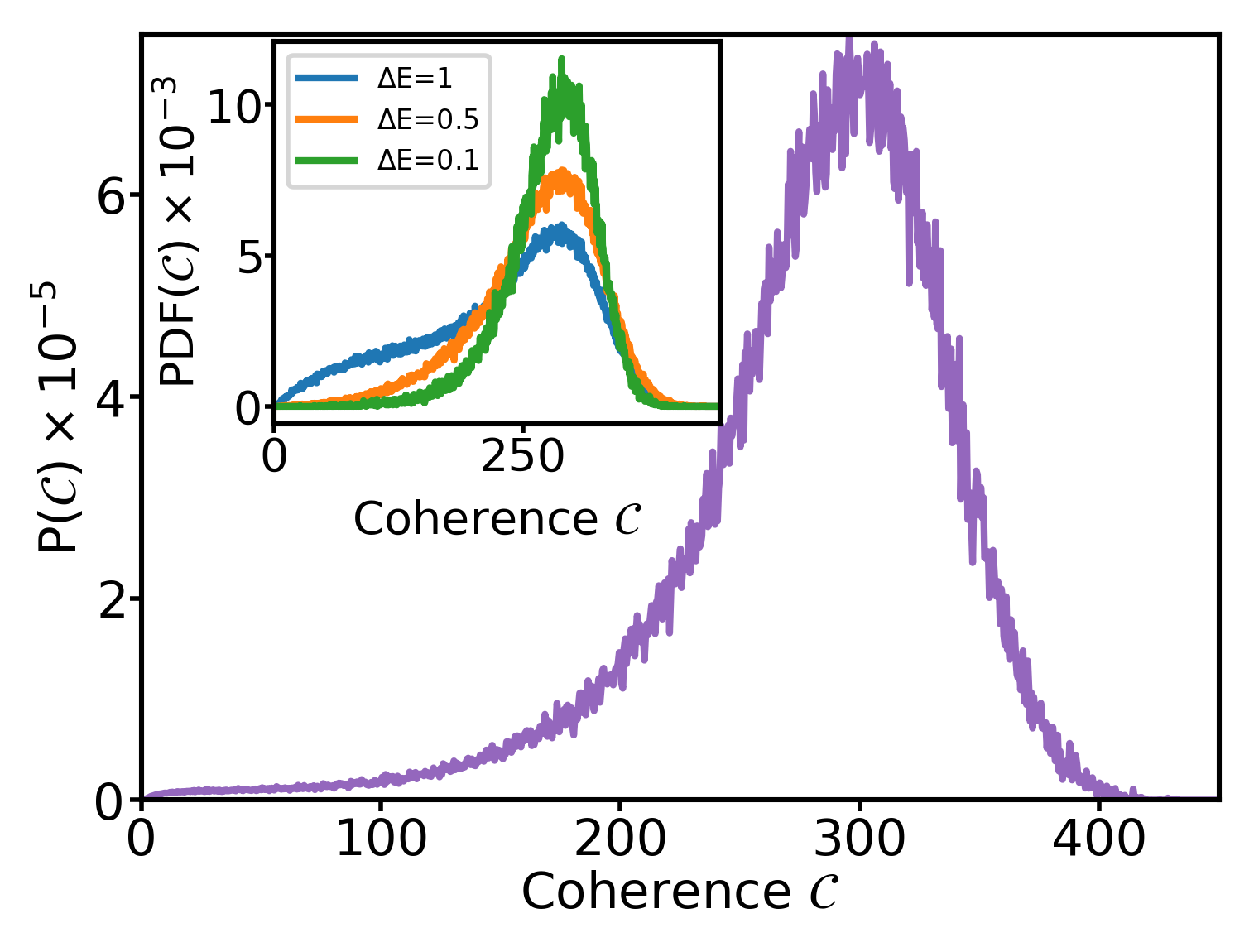}
\caption{\label{fig:absorp}
Probability to find states with coherence values $\mathcal{C}$ after exciting a gas of  $N=1000$ Rydberg atoms in the $\down$ state with a microwave pulse (no blockade). The following parameters are used:  We choose the $50s$ and $50p$ states of rubidium at a density around $10^{8}$cm$^{-3}$ so that the relevant interactions are on the order of $V_0\approx 10\, \mathrm{ MHz}$.
We use a rectangular microwave pulse with duration $t_{\rm mw}=500\,\mathrm{ns}$, a carrier frequency resonant with the transition frequency $\omega_{ps}$, and an electric field strength of $F_{\rm mw}\approx 5 \cdot 10^{-16}\mathrm{ au}$.  
The curve shown in Fig.~\ref{fig:absorp} stems from averaging  $\sim 1000$ realizations.
The inset shows marginal coherence distributions obtained from Fig.~\ref{fig:coherence_eigen_3D} where only energy intervals [-$\Delta E$, +$\Delta E$] around $E = 0$, rather than  the entire range as in  Fig.~\ref{fig:coherence_eigen_3D}, are considered. Approximately $0.5N$ states lie in this energy inteveral when $\Delta E = 1$; this fraction decreases to $0.32N$ for $\Delta E = 0.5$ and $0.084N$ for $\Delta E = 0.1$. 
}
\end{figure}

One practical issue is how to access these delocalized states.
Note that the random nature of the gas implies that we do not know the arrangement of the atoms, and equivalently the exact eigenenergies of the delocalized states. Additionally, we found that most of the delocalized states have only small oscillator stengths \cite{ourJPB}.  Nevertheless, short microwave pulses allow for selective excitation of states with large coherence.
This is demonstrated in Fig.~\ref{fig:absorp}, where we show the probability to find states with coherence $\mathcal{C}$ in a gas of 1000 atoms. 
Essentially, only states with coherence larger than 200 are populated.
To obtain this {curve}, we have solved the time-dependent Schr\"odinger equation taking the initial state $\ket{G}$, the single-exciton states $\ket{n}$, and the coupling to the microwave explicitly into account. 
Although, we have chosen parameters for which multi-exciton states can be ignored, the probability to have excited states remains nevertheless sufficiently large to be probed in experiment, being about 1\%. 
For smaller numbers of $N$ (up to $N=200$) we  have performed full calculations taking the two-exciton states into account. From these results we confirmed that our choice of parameters ensures negligible population of the two-exciton states. 
We note that it also interesting to excite and investigate the multi-exciton states, but this goes beyond the present work.
The basic reason why this simple excitation scheme works can be understood by considering the coherence distribution for energies around $E=0$, corresponding to the transition energy {$\omega_{sp}$}, in Fig.~\ref{fig:coherence_eigen_3D}. 
It is evident that the states at this energy have only large coherences (see inset of Fig.~\ref{fig:absorp}).
By choosing a microwave pulse that is weak enough to only couple to states with $E\approx 0$ and long enough such that the spectral width is also small, only these states are populated. 
We note that for each single shot (i.e.\ realization of the gas and microwave pulse) one will typically be in a coherent superposition of a few delocalized eigenstates.

In conclusion, we have undertaken an extensive numerical study of the properties of the collective eigenstates of an excitation in a random medium with long-range interactions. 
We stress that our observations are generic to a variety of physical situations with long-range interactions between randomly placed particles, although the random Rydberg gas emphasized here, having naturally long-range interactions with rich angular structure, random statistics, and the mechanism of Rydberg blockade for eliminating localized states, is an ideal physical realization. 
As demonstrated by Figs.~\ref{fig:intro} and~\ref{fig:blockade}, we find that the majority of eigenstates in a random gas are highly delocalized, with coherences extending upwards of one-third of the atoms.
There is also a clear asymmetry in this distribution with respect to the mean eigenenergy which is barely visible in the marginal distribution. We have studied 3D and 2D arrangements with different power-laws and different angle-dependencies in the interaction and found that a complicated picture emerges \cite{ourJPB}. {This} indicates a sensitive dependence on dimension and anisotropy of the interaction which deserves further detailed studies. 
We have shown that a promising way to reach the delocalized eigenstates is by using microwave pulses that are short compared to typical Rydberg lifetimes and the time-scales of dipole-dipole induced motion.
 While for the strongly interacting dimer states the interaction potential leads to quite fast atomic motion   \cite{LiTaJa06_27_} for the extended states we do not expect fast motion, since the interaction is smaller than in the dimer states and the induced forces are further reduced by the delocalization \cite{AtEiRo08_045030_}.  
An interesting perspective is to study the resulting adiabatic and non-adiabatic dynamics of such extended states \cite{WEiRo11_153002_}.

The observation that there exist strongly delocalized states with appreciable oscillator strength (of order unity) may aid in the interpretation and understanding of the phase modulation spectroscopy of very dilute gases interacting through the resonant dipole potential, although in a totally different energetic regime as these were not Rydberg atoms. In such experiments unexpectedly large signals have been observed \cite{BrBiSt15_053412_} and, in the absence of a more compelling explanation, attributed to many body effects \cite{LiBrSt17_052509_,BrEiBa19_-_}. 
The delocalized states that we find here can greatly amplify such signals. Although a full explanation requires a study of the two or more exciton system, preliminary studies indicate that the 2-exciton states have a coherence length that  scales as  $N^2/4$. The Rydberg parameter range explored here allows one to perform similar experiments under a more controlled setting to try to unravel this puzzle.

\begin{acknowledgments}
We acknowledge funding from the DFG: grant EI 872/4-1
through the Priority Programme SPP 1929 (GiRyd).
AE acknowledges support from the DFG via a Heisenberg fellowship (Grant No EI 872/5-1). MTE acknowledges support from the Max-Planck Gesellschaft via the MPI-PKS visitors program and from an Alexander von Humboldt Stiftung postdoctoral fellowship.
\end{acknowledgments}

\bibliographystyle{apsrev4-1}

\end{document}